\documentclass[doublecol]{epl2}

\title{Thermostatistics of $\mu$-deformed analog of Bose gas model}
\shorttitle{Thermostatistics of $\mu$-Bose gas model} 

\author{A.M. Gavrilik\footnote{E-mail: omgavr@bitp.kiev.ua} \and I.I. Kachurik \and A.P. Rebesh}
\shortauthor{A.M. Gavrilik \etal}

\institute{
  \inst{} Bogolyubov Institute for Theoretical Physics - 14-b, Metrolohichna str.
 Kiev, 03680, Ukraine
}


\abstract{For the recently introduced $\mu$-deformed analog of Bose
gas model  ($\mu$-Bose gas model) we study some thermodynamical
aspects. Namely, we calculate total number of particles and, from
it, the deformed partition function, both involving dependence on
the deformation parameter $\mu$. Such dependence of thermodynamic
functions on the $\mu$-parameter is at the core of modification of
Bose gas model and arises through the use of new techniques given by
us, the $\mu$-calculus, an alternative to the well-known
$q$-calculus (Jackson derivative and integral). Necessary elements
of $\mu$-calculus are first presented. Then, for high temperatures
we obtain virial expansion of the equation of state and find five
first virial coefficients, as functions of $\mu$. At the other end,
for low temperatures the critical temperature of condensation
$T^{(\mu)}_c$ depending on $\mu$ is found and compared with the
usual $T_c$, and with the $T_c^{(p,q)}$ of earlier studied
$p,q$-Bose gas model. The internal energy, specific heat and the
entropy of $\mu$-Bose gas are also given, both for high and low
temperatures. Features peculiar for the $\mu$-Bose gas model are
emphasized.}

\begin{document}

\maketitle

\section{Introduction}
During last two decades diverse so-called deformed Bose gas models
have  appeared, to list some
\cite{Martin91,Manko,Chaichian,Monteiro,DaoudKibler,ShuChen02,AdGavr,GavrSigma,AlginFibOsc,ScarfoneSwamy09,GavrRebIntercepts,GR-12}.
Usually such deformations are based on respective deformed
oscillator models of which best known and best studied are the
$q$-oscillators of Arik-Coon \cite{ArikCoon76} and
Biedenharn-Macfarlane \cite{Biedenharn} as well as the 2-parameter
$p,q$-deformed or Fibonacci oscillators \cite{Chakrabarti91}. The
Tamm-Dancoff deformed oscillator \cite{Odaka} is less known one,
like a plenty of nonstandard one-parameter $q$-oscillators
\cite{Plethora08}, polynomially deformed ones \cite{Polynomially}
along with so-called quasi-Fibonacci oscillators \cite{Kachurik}, to
which the $\mu$-deformed oscillator \cite{Jann} does also belong.
These nonstandard deformed oscillator models are peculiar due to
some new, unusual properties, e.g. various energy level
degeneracies, nontrivial recurrency relations for energy spectra
etc. Those features make them plausible for application in diverse
fields of quantum physics.

Physical meaning of deformation per se, and of the involved
deformation  parameter(s) depends on the particular application of
deformed oscillator or deformed Bose gas model to concrete physical
system. When applying the model of ideal gas of {\it deformed}
bosons to computing the intercepts of the momentum correlation
functions \cite{AdGavr,GavrSigma,GavrRebIntercepts}, one can
effectively take into account the structure (compositeness) of
particles, or their finite volume, see \cite{GKM2011}, \cite{av95}.
In fact, comparison of theory with experimental data \cite{Abelev}
showing non-Bose type behavior of the two-pion correlation function
intercepts stresses the efficiency \cite{GavrSigma,AGP} of using
deformed versions of Bose gas model. Let us also mention successful
application of $q$-Bose gas setup to overcome the difficulty with
unstable phonon spectrum, confirmed \cite{Monteiro} by experimental
measurements of phonon lifetime in scattering of neutrons.

In the deformed analogs of Bose gas model, and the $\mu$-Bose gas
model in particular, one counts for  modified interparticle
interaction \cite{GR-12,AlginSenay} of quantum statistical origin.
The deformation may also absorb \cite{ScarfoneSwamy09} an
interaction present in the initially non-deformed system.

The $\mu$-Bose gas model was proposed in \cite{GavrRebIntercepts}
where (the intercepts of) the momentum correlation functions of 2nd,
3rd,  and $r$th order were obtained in closed form. On the other
hand, the $p,q$-deformed model was explored more extensively, see
e.g. \cite{AdGavr,AlginFibOsc,GR-12}. Say, besides similar $r$th
order correlation function intercepts explicitly derived in
\cite{AdGavr}, the thermodynamic aspects including total number of
particles, (virial expansion of) the equation of state with virial
coefficients, and the modified ($p,q$-deformed) critical temperature
for $p,q$-Bose gas, have been obtained \cite{GR-12}. What concerns
$\mu$-Bose gas, its thermodynamical quantities were not studied
before, and in this paper we make some steps to fill the gap.
Dealing with the $\mu$-Bose gas model we focus our attention on
thermodynamics. In Sec. 2 we recall necessary facts concerning the
$\mu$-deformed analog of Bose gas and devote a subsection to
elements of $\mu$-calculus. Then we explore basic thermodynamic
quantities. In Sec. 3 we derive the expression for the total number
of particles and, using that, the partition function (the both
quantities carry explicit $\mu$-dependence). Next, we study two
opposite regimes: for high temperature and low density we obtain
virial coefficients of the expanded equation of state using deformed
calculus ($\mu$-derivative); then, for the regime of low temperature
and high density the critical temperature of condensation is
obtained and its dependence on the deformation parameter $\mu$
studied. Few other thermodynamical functions are treated in Subsec.
3.3. The paper is ended with conclusions.

\section{Deformed analogs of Bose gas model}
Like in many papers studying deformed oscillators, see e.g.
\cite{Monteiro,ScarfoneSwamy09,AlginFibOsc} we deal in fact with the
(system of) related deformed bosons. One of the virtues of such
deformation is its ability to provide effective account of the
interaction between particles, their non-zero volume, or their inner
(composite) structure.

To study deformed analog of Bose gas model, namely the $\mu$-Bose
gas  as the model describing the system of deformed bosons, we work
the Hamiltonian\footnote{We use here symbol $\widetilde{\mu}$ for
the chemical potential in order to distinguish it from the
deformation parameter $\mu$; throughout below $\widetilde{\mu}$ will
be hidden in the fugacity, see  relation (\ref{eq.3}) below.}
\begin{equation}\label{eq.1}
H=\sum_i(\varepsilon_i-\widetilde{\mu})N_i
\end{equation}
Here $\varepsilon_i$ denotes kinetic energy of particle in the state
"$i$",  $N_i$ the particle number (occupation number) operator
corresponding to state "$i$". Similarly to the case of ordinary
bosons, we proceed as follow. Calculation of the thermodynamic
functions in the new model needs special mathematical tools
($\mu$-calculus) that will be presented below.

The special version of deformed Bose gas model termed $\mu$-Bose gas
model associated with $\mu$-deformed oscillators \cite{Jann} was
introduced and studied in \cite{GavrRebIntercepts}. Therein, and in
this paper, the thermal average of the operator $\mathcal{O}$  is
determined by the familiar formula
\begin{equation}\label{eq.2}
\langle \mathcal{O} \rangle=\frac{Tr(\mathcal{O}e^{-\beta H})}{Z},
\end{equation}
$Z$ being the grand canonical partition function.
 This function is (the logarithm of it)
\begin{equation}\label{eq.3}
\ln Z=-\sum_i\ln(1-ze^{-\beta\varepsilon_i})
\end{equation}
with the fugacity $z=e^{\beta\widetilde{\mu}}$.
 The usual (non-deformed) formula for the total number
of particles i.e.
\begin{equation}\label{eq.4}
N=z\frac{d}{dz}\ln Z,
\end{equation}
in our treatment will be modified, see Sec. 3.

\subsection{Jackson derivative and its $p,q$-extensions}
To derive thermodynamical functions for a deformed analog of Bose
gas  model, one needs an extension of usual treatment, in particular
what concerns derivatives. Say, instead of usual derivative one uses
the Jackson or $q$-derivative \cite{Kac}
 \begin{equation}\label{eq.5}
\Bigl(\frac{d}{dx}\Bigr)_qf(x)\equiv  \mathcal{D}^{(q)}_x
f(x)=\frac{f(qx)-f(x)}{qx-x} .
\end{equation}
Its consistency requires that at $q\rightarrow 1$ we recover $d/dx$:
\begin{equation}\label{eq.6}
\mathcal{D}_x^{(q)}\stackrel{q\rightarrow1}{\longrightarrow}\frac{d}{dx}.
\end{equation}
When acting on the monomial in $x$ the $q$-derivative gives
\begin{equation}\label{eq.7}
\mathcal{D}^{(q)}_x x^n=\frac{(qx)^n-x^n}{qx-x}=[n]_qx^{n-1},  \quad
[n]_q\equiv\frac{q^n-1}{q-1},
\end{equation}
with $[n]_q$ the $q$-bracket. At $q\rightarrow 1$,
$\displaystyle\lim_{q\rightarrow1}[n]_q=n$ and the ordinary
derivative is regained. With respect to $x\mathcal{D}^{(q)}_x$, the
monomials behave as eigenvectors.

A $p,q$-extension of Jackson derivative is also known, see e.g.~\cite{Floreanini93}.
This is the operation $\mathcal{D}^{(p,q)}_x$ acting as
\begin{equation}\label{eq.8}
\mathcal{D}^{(p,q)}_xf(x)=\frac{f(px)-f(qx)}{px-qx},
\end{equation}
$$
 \quad \mathcal{D}^{(p,q)}_xx^n=[n]_{p,q}x^{n-1}, \quad [n]_{p,q}\equiv\frac{p^n-q^n}{p-q}.
$$
For $p=1$ we recover the Jackson derivative $\mathcal{D}^{(q)}_x $ in (\ref{eq.5}), (\ref{eq.7}).
Remark that somewhat modified version $\widetilde{\mathcal{D}}_z^{(p,q)}$
of the $p,q$-extended Jackson derivative was used in \cite{AlginFibOsc}.

\subsection{Elements of $\mu$-calculus}
The $\mu$-derivative is an alternative to Jackson derivative.
For the needs of this paper ($\mu$-Bose gas model), in some analogy with the above
extensions of $d/dx$ we introduce the $\mu$-deformed derivative:
\begin{equation}\label{eq.10}
\mathcal{D}^{(\mu)}_x x^n\!=\![n]_{\mu}x^{n\!-\!1},  \quad
[n]_{\mu}\!\equiv\!\frac{n}{1+\mu n} \quad \mbox{($\mu$-bracket)}.
\end{equation}
This particular form of $\mu$-derivative is linked to the
$\mu$-bracket appearing  in the study of $\mu$-deformed (Jannussis)
oscillator \cite{Jann}. At $\mu\rightarrow 0$ the $\mu$-extension
$\mathcal{D}^{(\mu)}_x$ goes over directly into the usual derivative
$d/dx$.

Although knowledge of the action on monomials $x^m$ for the
$\mu$-derivative  is enough for the goals of our treatment, let us
also indicate how the $\mu$-derivative can be defined when operating
upon generic function $f(x)$:
\begin{equation}\label{eq.11}
\mathcal{D}^{(\mu)}_xf(x)=\int^1_0dtf'_x(t^{\mu}x),  \qquad
f'_x(t^{\mu}x)=\frac{d f(t^{\mu}x)}{d x}.
\end{equation}
As seen, formula (\ref{eq.10}) stems from this general definition.

The $k$th power of $\mu$-derivative on a monomial $x^n$ yields
\begin{equation}\label{eq.12}
(\mathcal{D}^{(\mu)}_x)^kx^n=\frac{[n]_{\mu}!}{[n-k]_{\mu}!}x^{n-k},
\qquad [n]_{\mu}!\equiv\frac{n!}{(n;\mu)} ,
\end{equation}
where $(n;\mu)\equiv(1+\mu)(1+2\mu)...(1+n\mu)$.

Generalization of $\mu$-derivative (\ref{eq.11}) to its $q,{\mu}$-
or $(p,q;\mu)$-deformed extensions can be obtained if instead of
$(d/dx)f(t^{\mu}x)$ in (\ref{eq.11}) we take resp.
$\mathcal{D}^{q}_xf(t^{\mu}x)$ or
$\mathcal{D}^{(p,q)}_xf(t^{\mu}x)$. These extended cases correspond
to the $(q;\mu)$- or $(p,q;\mu)$-deformed quasi-Fibonacci
oscillators treated in \cite{Kachurik}.

The inverse $\bigl({\mathcal{D}^{(\mu)}_x}\bigr)^{-1}$ or
antiderivative of  the $\mu$-derivative $D^{(\mu)}_x$ in
(\ref{eq.10}) and (\ref{eq.11}) can be also defined, though we do
not give it here, since it will not be used below.

So, to develop thermodynamics of $\mu$-Bose gas model we apply, at
proper point, the modified derivative $D^{(\mu)}_z$ instead of usual
$d/dz$. As result, through the $\mu$-analog of derivative the
deformation parameter gets involved in the treatment; the system
becomes $\mu$-deformed. It is essential that, at small values of
$\mu$, both the usual and the deformed  
derivative of a function have similar behavior. This can be verified
by acting with the deformed and usual types of derivative on the
monomial, logarithmic, exponential function, and others. Such
property of $\mu$-derivative justifies (at least partly) its very
use in calculating thermodynamical quantities of $\mu$-Bose gas.

By the use of $\mu$-bracket $[n]_{\mu}$ and $\mu$-factorial
$[n]_{\mu}!$, see (\ref{eq.10}), (\ref{eq.12}), we earn
$\mu$-deformed analogs of elementary functions: $\mu$-exponential
$exp_{\mu}(x)$, $\mu$-logarithm $\ln_{\mu}(x)$ (where the
$\mu$-numbers $[n]_{\mu}$ and $\mu$-factorial
$[n]_{\mu}!=[n]_{\mu}[n-1]_{\mu}...[2]_{\mu}[1]_{\mu}$ appeared).
Certain special functions, namely $\mu$-analog of polylogarithms,
will appear in Sec. 3.

\subsection{$D^{(\mu)}_x$ acting on product of functions
 or $\mu$-analog of Leibnitz rule}
Consider the rule of $\mu$-differentiation when acting on the
product $f(x)\cdot g(x)$ of $f(x)$ and $g(x)$. From definition $(10)$, in the
case of monomials $f(x)=x^n$ and $g(x)=x^m$ we have the
relation
\begin{equation}\label{eq.13}
D^{(\mu)}_x \left( x^n x^m\right) = D^{(\mu)}_x \left( x^m
x^n\right) = \frac{n+m}{1+\mu(n+m)}\, x^{n+m-1} .
\end{equation}
General formula for $D^{(\mu)}_x$ acting on $f(x)\cdot g(x)$ does also
exist but we do not dwell on it here.

\section{Thermodynamics of $\mu$-Bose gas model}

Now let us study thermodynamics of $\mu$-deformed analog of
Bose gas using the elements of $\mu$-calculus given above. For
the gas of non-relativistic particles the both regimes of high and low
temperatures will be treated in what follows.

\subsection{Total number of particles}
In the Bose gas model the known relation giving total number of particles is
\begin{equation}\label{eq.14}
N=z\frac{d}{dz}\ln Z.
\end{equation}
To develop thermodynamics of $\mu$-analog of Bose gas model, this
formula for total number $N$ is to be modified and we adopt
\begin{equation}\label{eq.15}
N\equiv N^{(\mu)}=z\mathcal{D}^{(\mu)}_z\ln Z
=-z\mathcal{D}^{(\mu)}_z\sum_i\ln(1-ze^{-\beta\varepsilon_i}) ,
\end{equation}
where $\mathcal{D}^{(\mu)}$ is the $\mu$-derivative from
(\ref{eq.10}). For $\mu\geq 0$, we apply this to the $\log$ of
partition function in (\ref{eq.3}) to get
\begin{equation}\label{eq.16}
N^{(\mu)}\!=\!z\sum_i\sum_{n=1}^{\infty}\frac{e^{-\beta\varepsilon_in}}{n}[n]_{\mu}z^{n-1}\!=\!
\sum_i\sum_{n=1}^{\infty}\frac{[n]_{\mu}}{n}(e^{-\beta\varepsilon_i})^nz^n.
\end{equation}
Note there should be $0\leq|ze^{-\beta\varepsilon_i}|<1$ in (\ref{eq.16}).
As we deal with non-relativistic particles, the energy $\varepsilon_i$ is taken as
\begin{equation}\label{eq.17}
\varepsilon_i=\frac{\overrightarrow{p}_i\overrightarrow{p}_i}{2m}
=\frac{|p|^2}{2m}=\frac{p_i^2}{2m}.
\end{equation}
Here $\overrightarrow{p}_i$ is the 3-momentum of particle in $i$-th
state and  $m$ the particle mass.

Clearly, at $z\rightarrow 1$ the expression under summation symbol
in (\ref{eq.16}) diverges when $p_i=0, i=0$. In next subsection we
will assume the $i=0$ ground state to be associated with a
macroscopically large occupation number. Here, even though $z\neq
1$, we nevertheless separate the term with $p_i=0$ from the
remaining sum:
\begin{equation}\label{eq.18}
N^{(\mu)}={\sum_i} ' \sum_{n=1}^{\infty}\frac{[n]_{\mu}}{n}(e^{-\beta\varepsilon_i})^nz^n +
\sum_{n=1}^{\infty}\frac{[n]_{\mu}}{n}z^n.
\end{equation}
The ``prime'' of the sum symbol in (\ref{eq.18}) means that the
$i=0$ term is dropped from the sum. For large volume $V$ and large
$N$ the spectrum of single-particle states is almost continuous so
we replace the sum in (\ref{eq.16}) by integral:
\begin{equation}\label{eq.19}
\sum_i\rightarrow \frac{V}{(2\pi \hbar)^3}\int d^3k.
\end{equation}
That is, we isolate the ground state and include the contribution
from all other states in the integral. Now, to compute the total
number of particles we perform integration  over 3-momenta using
spherical coordinates:
 \begin{equation}\label{eq.20}
  N^{(\mu)}\!=\!\frac{4\pi V}{(2\pi\hbar^2)^3}\!\sum_{n=1}^{\infty}\!\frac{[n]_{\mu}z^n}{n}\!\!\int^{\infty}_0 \!\!\!\!p^2e^{-\frac{\beta p^2}{2m}}dp +\!\sum_{n=1}^{\infty}\!\frac{[n]_{\mu}z^n}{n}.
 \end{equation}
The lower limit of the integral can still be taken as zero, because the ground state,
$p_0$, does not contribute to the integral anyway.
Then, after performing the last integration by parts
we obtain for the ($\mu$-deformed, i.e. depending on $\mu$)
total number of particles the expression
\begin{equation}\label{eq.21}
N^{(\mu)}=\frac{V}{\lambda^3}\sum_{n=1}^{\infty}\frac{[n]_{\mu}}{n^{5/2}}z^n+
N_0^{(\mu)}, \quad N_0^{(\mu)}\equiv
\sum_{n=1}^{\infty}\frac{[n]_{\mu}}{n}z^n ,
\end{equation}
where $\lambda=\sqrt{\frac{2\pi \hbar^2}{mkT}}$ is the thermal wavelength.
 The expression in (\ref{eq.21})
can be rewritten using the $\mu$-analog of Bose-Einstein function. As the
result, for $N^{(\mu)}$ we obtain
\begin{equation}\label{eq.22}
N^{(\mu)}=\frac{V}{\lambda^3}g_{3/2}^{(\mu)}(z)+g_0^{(\mu)}(z),
\qquad  g_0^{(\mu)}(z)=N_0^{(\mu)},
\end{equation}
where $g_0^{(\mu)}(z)$ is a particular case of the following
$\mu$-polylogarithm ($\mu$-analog of well known Bose-Einstein function
or polylogarithm $g_l(z)=\sum_{n=1}^{\infty}z^n/n^l$):
\begin{equation}\label{eq.23}
g_{l}^{(\mu)}(z)=\sum_{n=1}^{\infty}\frac{[n]_{\mu}}{n^{l+1}}z^n.
\end{equation}
At $\mu\rightarrow 0$, from this the usual $g_l(z)$ function is recovered.

One can show that the positive real parameter $\mu$ does not influence the
convergence region and, like for the ordinary
$g$-function $g_l(z)$, there should be $|z|<1$. This is in contrast with the
restriction $z<1/q$ in Ref.~\cite{ShuChen02}
valid for the version of $q$-Bose gas model considered therein.

It is useful to rewrite the expression in (\ref{eq.22}) for total number of particles as
\begin{equation}\label{eq.24}
\frac{1}{v}=\frac{1}{\lambda^3}g^{(\mu)}_{3/2}+\frac{N_0^{(\mu)}}{V},
\qquad v\equiv\frac{V}{N^{(\mu)}}.
\end{equation}
This will be exploited in subsequent analysis.

\subsection{Deformed grand partition function}
We assume that all the known relations between thermodynamical
functions in case of usual Bose gas thermodynamics are shared by
the $\mu$-deformed analog of Bose gas model. This
means that well-known relations are formally the same both for usual
Bose gas model and its $\mu$-deformed counterpart. The only thing
that should be stressed is that for $\mu$-Bose gas model all the
thermodynamical functions are $\mu$-dependent.

So, to get deformed partition function $\ln Z^{(\mu)}$ we take
\begin{equation}\label{eq.25}
N^{(\mu)}=z\frac{d}{dz}\ln Z^{(\mu)}
\end{equation}
and invert it, that is,
\begin{equation}\label{eq.26}
\ln Z^{(\mu)}=\Bigl(z\frac{d}{dz}\Bigr)^{-1}N^{(\mu)}.
\end{equation}
To perform the operation $\bigl(z\frac{d}{dz}\bigr)^{-1}$ to get
$\ln Z^{(\mu)}$, one may either integrate: $\ln
Z^{(\mu)}=\int\!dz~z^{-1}N^{(\mu)}$ or merely apply the
following property, valid on the monomials $z^k$, for any function
$f(z\frac{d}{dz})$ with power series expansion:
\begin{equation}\label{eq.27}
f\Bigl(z\frac{d}{dz}\Bigr)z^k=f(k)z^k.
\end{equation}
With account of this, from (\ref{eq.26}), (\ref{eq.27}) and
(\ref{eq.18})-(\ref{eq.21}) we draw:
$$
\ln Z^{(\mu)}=\Bigl(z\frac{d}{dz}\Bigr)^{-1}\left(\frac{V}{\lambda^3}\sum_{n=1}^{\infty}\frac{[n]_{\mu}}{n^{5/2}}z^n+
\sum_{n=1}^{\infty}\frac{[n]_{\mu}}{n}z^n\right)=
$$
$$
=\frac{V}{\lambda^3}\sum_{n=1}^{\infty}\frac{[n]_{\mu}}{n^{5/2}}\Bigl(z\frac{d}{dz}
\Bigr)^{-1}z^n+\sum_{n=1}^{\infty}\frac{[n]_{\mu}}{n}\Bigl(z\frac{d}{dz}\Bigr)^{-1}z^n=
$$
\begin{equation}\label{eq.28}
=\frac{V}{\lambda^3}\sum_{n=1}^{\infty}\frac{[n]_{\mu}}{n^{5/2}}(n)^{-1}z^n+\sum_{n=1}^{\infty}\frac{[n]_{\mu}}{n}(n)^{-1}z^n.
\end{equation}
The latter result can be written as (see definition (\ref{eq.23}))
\begin{equation}\label{eq.29}
\ln Z^{(\mu)}=\frac{V}{\lambda^3}g^{(\mu)}_{5/2}+g^{(\mu)}_1
\end{equation}
or as
\begin{equation}\label{eq.30}
Z^{(\mu)}(z,T,V)=\exp\left(\frac{V}{\lambda^3}g^{(\mu)}_{5/2}(z)+g^{(\mu)}_1(z)\right).
\end{equation}
Formulas (\ref{eq.28})-(\ref{eq.30}) constitute
our basic result for $\mu$-deformed partition
function\footnote{Recall that the same result follows by
integrating over $dz$ the expression $z^{-1}N^{(\mu)}$, see eq.
(\ref{eq.25}).}. Using (\ref{eq.30}) it is possible to
derive other thermodynamical functions or relations as well. In particular,
the equation of state reads:
\begin{equation}\label{eq.31}
\frac{PV}{kT}=\ln Z^{(\mu)}=\frac{V}{\lambda^3}g^{(\mu)}_{5/2}(z)+g^{(\mu)}_1(z).
\end{equation}

Remark that a different path of deriving the deformed $\mu$-depended
partition function could be the usage of one-particle $\mu$-deformed
distribution function  $\langle n\rangle^{(\mu)}$ whose exact
analytical expression was found in Ref.~\cite{GavrRebIntercepts}. It
can be shown that the result derived in such way will be slightly
different from that given in (\ref{eq.30}). That means in
(\ref{eq.28}) or (\ref{eq.30}) a different $\langle
n\rangle^{(\mu)}$ is implicit.

\subsection{Virial expansion of the equation of state and virial coefficients}
Consider the regime of high temperature and low density
$\lambda^3/v\ll 1$. As explained in handbooks \cite{Pathria}, in
this case the second term in  (\ref{eq.24}) is negligibly small and
likewise in eq.~(\ref{eq.31}). The expression (\ref{eq.24}) and the
equation of state (\ref{eq.31}) take the form
\begin{equation}\label{eq.32}
\frac{1}{v}=\frac{1}{\lambda^3}g^{(\mu)}_{3/2}(z), \qquad \frac{Pv}{kT}=\frac{v}{\lambda^3}g^{(\mu)}_{5/2}(z).
\end{equation}
From the first equality in (\ref{eq.32}) we have
\begin{equation}\label{eq.33}
\frac{\lambda^3}{v}=z+\frac{[2]_{\mu}}{2^{5/2}}z^2+\frac{[3]_{\mu}}{3^{5/2}}z^3+
\frac{[4]_{\mu}}{4^{5/2}}z^4+\frac{[5]_{\mu}}{5^{5/2}}z^5+...,
\end{equation}
with $[n]_{\mu}$ the number, see (\ref{eq.10}).
Inverting the series in (\ref{eq.33}) we derive virial expansion for the equation of state
\begin{equation}\label{eq.34}
\frac{Pv}{kT}=1+A\Bigl(\frac{\lambda^3}{v}\Bigr)+B\Bigl(\frac{\lambda^3}{v}\Bigr)^2+
C\Bigl(\frac{\lambda^3}{v}\Bigr)^3+D\Bigl(\frac{\lambda^3}{v}\Bigr)^4+...
\end{equation}
where the explicit form of virial coefficients is
\begin{equation}\label{eq.35}
A=-\frac{[2]_{\mu}}{2^{7/2}[1]^2_{\mu}}, \qquad B=\frac{[2]_{\mu}^2}{2^{5}[1]_{\mu}^4}-\frac{2[3]_{\mu}}{3^{7/2}[1]_{\mu}^3},
\end{equation}
$$
C=-\frac{5[2]^3_{\mu}}{2^{17/2}[1]^6_{\mu}}+\frac{[2]_{\mu}[3]_{\mu}}{2^{5/2}3^{3/2}[1]^5_{\mu}}-\frac{3[4]_{\mu}}{2^{7}[1]^4_{\mu}},
$$
$$
D=-\frac{[2]^2_{\mu}[3]_{\mu}}{2^33^{3/2}[1]^7_{\mu}}+\frac{2[3]^2_{\mu}}{3^{5}[1]^6_{\mu}}+
\frac{7[2]^4_{\mu}}{2^{10}[1]^8_{\mu}}-\frac{4[5]_{\mu}}{5^{7/2}[1]^5_{\mu}}+\frac{[2]_{\mu}[4]_{\mu}}{2^{11/2}[1]^6_{\mu}}.
$$
As seen the deformation parameter appears in all the virial coefficients in
specific manner, through the $\mu$-integers.
 Here we encounter a very unusual feature: the appearance of (powers of)
the $\mu$-unity $[1]_{\mu}$. While its analog was ``hidden'' in the
virial coefficients of $q$-Bose or $p,q$-Bose gases\footnote{Let us
note that in \cite{GR-12}, in the expression for the virial
coefficient $D$ the  (corrected) second term should be
$-[2]^2_{p,q}[3]_{p,q}2^{-3}3^{-3/2}$, compare with the first term
in  (\ref{eq.35}).}, because of equality $[1]_{p,q}=1$, see
\cite{GR-12}, here due to $[1]_{\mu}\neq 1$, the $\mu$-deformed
virial coefficients contain $[1]_{\mu}$ squared and higher powers of
it.

Physical meaning of the parameter $\mu$ may be commented as follows.
Virial coefficients in case of usual Bose gas model, as known,
reflect  the effective (2-particle, 3-particle etc.) interaction of
the quantum correlation or quantum statistics origin. In $\mu$-Bose
gas, besides that, the inner structure (compositeness) of particles
can effectively be taken into account by means of deformation and
the $\mu$ parameter; clearly, all that adds some extra amount of
effective interaction.   In effect, by changing the value of
deformation parameter we can control/regulate both the value, and
even the sign of virial coefficient(s) and thus can even gain
repulsive instead of attractive (or vice versa) effective
interparticle interaction in the model system. That is, by moving
the value of $\mu$ we can effectively change and thus control the
quantum statistics of particles (compare with other deformed Bose
gas models \cite{ScarfoneSwamy09,GR-12,AlginSenay,Ubriaco}).

\subsection{Critical temperature of condensation}
For the regime of low temperature and high density we obtain  (like
in \cite{GR-12} for the $p,q$-Bose gas) the critical temperature
$T^{(\mu)}_c$ of condensation in the considered $\mu$-deformed
analog of Bose gas model. We start with eq.~(\ref{eq.24}) and
rewrite it as
\begin{equation}\label{eq.36}
\frac{N_0^{(\mu)}}{V}=\frac{\lambda^3}{v}-g^{(\mu)}_{3/2}(z).
\end{equation}

\begin{figure}[width=7cm]
\onefigure{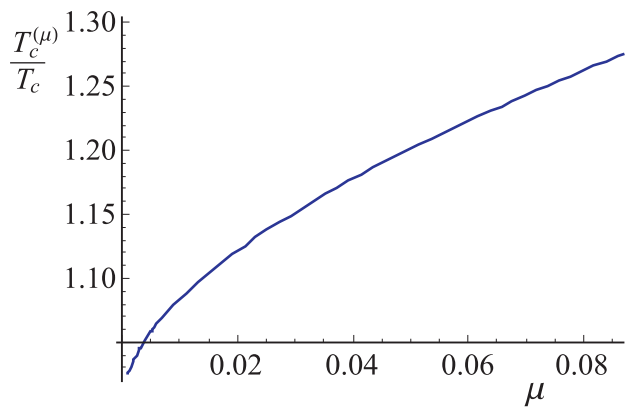} \vspace{-3mm} \caption{The ratio
$T_c^{(\mu)}/{T_c}$ versus deformation parameter $\mu$.}
 \label{fig.1}
\end{figure}

The critical temperature $T_c^{(\mu)}$ of $\mu$-Bose gas model is
determined from the equation $\lambda^3/{v}=g^{(\mu)}_{3/2}(1)$,
that gives:
\begin{equation}\label{eq.37}
T_c^{(\mu)}=\frac{2\pi\hbar^2/mk}{\bigl(vg^{(\mu)}_{3/2}(1)\bigr)^{2/3}}.
\end{equation}
From the latter we infer the ratio of critical temperature
$T_c^{(\mu)}$ and the critical temperature $T_c$ of usual Bose gas:
\begin{equation}\label{eq.38}
\frac{T_c^{(\mu)}}{T_c}=\Biggl(\frac{2.61}{g^{(\mu)}_{3/2}(1)}\Biggr)^{2/3}.
\end{equation}
Figure \ref{fig.1} shows how the ratio (\ref{eq.38}) depends on the
deformation parameter $\mu$.

We see that the ratio $T_c^{(\mu)}/{T_c}$, similar to the case of
$p,q$-Bose gas model \cite{GR-12}, has a very important feature: the
greater  is the extent of deformation (here measured by $\mu$) the
higher is $T_c^{(\mu)}$. In the no-deformation limit $\mu\rightarrow
0$ we have $T_c^{(\mu)}/{T_c}=1$, that is, the $\mu$-critical
temperature tends to usual one, $T_c^{(\mu)}\rightarrow T_c$ (a kind
of consistency).

\subsection{Other thermodynamical functions}
For the studied $\mu$-deformed analog of Bose gas model we take the
definition  of thermodynamical functions as in \cite{Pathria}.
 Then, the internal energy $U^{(\mu)}$ of $\mu$-Bose gas is found as
$U^{(\mu)}=-\left(\partial\ln Z^{(\mu)}/\partial\beta\right)_{z,V}$,
and we examine the both cases $T>T_c^{(\mu)}$ and $T\leq
T_c^{(\mu)}$:
\begin{equation}\label{eq.39}
\frac{U^{(\mu)}}{N^{(\mu)}}= \left\{
 \begin{array}{rl}
\displaystyle\frac{\mathstrut 3}{2}\frac{kTv}{\lambda^3} g^{(\mu)}_{5/2}(z), & T>T_c^{(\mu)} \vspace{2mm} \\
\displaystyle\frac{\mathstrut3}{2}\frac{kTv}{\lambda^3} g^{(\mu)}_{5/2}(1), & T\leq T_c^{(\mu)}.
\end{array}
 \right.
\end{equation}
Using the expressions for internal energy of $\mu$-Bose gas we easily
obtain the specific heat of the system using the relation
$C_v=\left(\partial U/\partial T \right)_{N,V}$, and the result is
\begin{equation}\label{eq.40}
\frac{C_v^{(\mu)}}{N^{(\mu)}k}=\! \left\{
 \begin{array}{rl}
\displaystyle\frac{\mathstrut 15}{4}\frac{v}{\lambda^3} g^{(\mu)}_{5/2}(z)-
\displaystyle\frac{\mathstrut 9}{4}\frac{g^{(\mu)}_{3/2}{(z)}}{g^{(\mu)}_{1/2}{(z)}}, & T>T_c^{(\mu)} \vspace{2mm} \\
\displaystyle\frac{\mathstrut 15}{4}\frac{v}{\lambda^3} g^{(\mu)}_{5/2}(1), & T\leq T_c^{(\mu)}.
 \end{array}
 \right.
\end{equation}
As seen, both in (\ref{eq.39}) and (\ref{eq.40}) the $\mu$-deformed
analog  ($\mu$-polylogarithm) of Bose function $g_n(z)$ does appear.
This fact may have interesting implications.

From the well-known relation for entropy $S=\ln Z+\beta
U-\beta\widetilde{\mu} N$, within the $\mu$-analog of Bose gas for
the regimes of high resp. low temperatures we deduce the expressions
\begin{equation}\label{eq.41}
    \frac{S^{(\mu)}}{N^{(\mu)}k}=\frac{5}{2}\frac{v}{\lambda^3}g^{(\mu)}_{5/2}(z)
    - \ln z,
    \qquad T>T^{(\mu)}_c,
\end{equation}
\begin{equation}\label{eq.42}
   \frac{S^{(\mu)}}{N^{(\mu)}k}=\frac{5}{2}\frac{v}{\lambda^3}g^{(\mu)}_{5/2}(1),
    \qquad T\leq T^{(\mu)}_c.
\end{equation}
Notice that from (\ref{eq.39}), (\ref{eq.40}) and (\ref{eq.42}) we
infer an  interesting proportionality valid for $T<T^{(\mu)}_c$:
$$
\frac{U^{(\mu)}}{T}=\frac{2}{5}C_v^{(\mu)}=\frac{3}{5}S^{(\mu)}
$$
(it is formally similar to the case of usual Bose gas).

\begin{figure}
\onefigure[width=8.5cm]{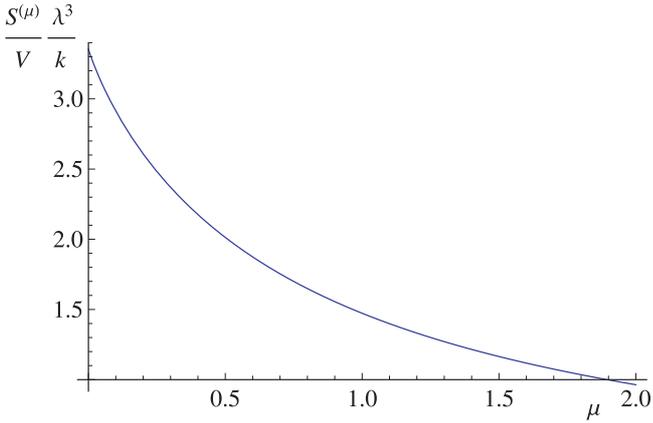} \vspace{-3mm} \caption{Ratio
$\frac{S^{(\mu)}}{V}$ (times $\frac{\lambda^3}{k}$) versus
deformation parameter $\mu$.}
 \label{fig.2}
\end{figure}

Let us visualize the dependence of $\frac{S^{(\mu)}}{V}$ in
(\ref{eq.42}) on $\mu$, see Fig. \ref{fig.2}.
Here we find a \underline{remarkable fact}: enhancing deformation,
by raising of $\mu$-values, results in falling entropy-per-volume
$\frac{S^{(\mu)}}{V}$ (in units of $\frac{k}{\lambda^3}$), from
$\approx 3.354$ at $\mu=0$ to zero in the large deformation
asymptotics. That is, {\it the larger deformation the lesser is
chaoticity} in the system.


\section{Conclusions}
For the $\mu$-deformed analog of Bose gas model proposed in
\cite{GavrRebIntercepts} in the present work we explored a number of
thermodynamic relations.  Having presented elements of
$\mu$-calculus, using the $\mu$-deformed analog of derivative, we
have calculated the total mean number of particles. This result
allowed us to explicitly derive the ($\mu$-deformed) partition
function. Due to usage of $\mu$-derivative the deformation parameter
has appeared in the expressions for total number of particles and
for grand partition function. In the formulas for $N^{(\mu)}$ and
for the $\log$ of $\mu$-partition function, the $\mu$-analogs
$g^{(\mu)}_{3/2}(z)$ and $g^{(\mu)}_{5/2}(z)$ of polylogarithms
$g_{n}(z)$ have naturally appeared.

For high temperature regime we have obtained explicit virial
expansion for the equation of state and five ($\mu$-dependent)
virial coefficients. A really surprising thing we encounter is the
remarkable appearance of powers of the $\mu$-unit
$[1]_{\mu}=\frac{1}{1+\mu}$ in $\mu$-dependent virial coefficients.
Say in $A$ in (\ref{eq.35}), because of square of $[1]_{\mu}$ in the
denominator, the increase of deformation leads to some extra raising
of strength of effective two-particle interaction. Similar features
concern higher virial coefficients $B$, $C$, $D$.

At $\mu\neq 0$ the deformed system principally differs from bosonic
one, and variation of $\mu$ smoothly changes the statistics of
particles. That is why, by controlling the $\mu$ value one can even
achieve switching of the sign of virial coefficient(s) that yields
drastic change of quantum statistics. At $\mu=0$ (no-deformation
limit), as it should, the obtained $\mu$-deformed virial
coefficients reduce to the known virial coefficients of usual Bose
gas.

For the low temperature regime the critical temperature of
condensation  depending explicitly on $\mu$ is obtained.
 The dependence of ratio $T_c^{(\mu)}/{T_c}$ on the parameter $\mu$ shows
that critical temperature in $\mu$-Bose gas is higher than critical
temperature $T_c$ of usual Bose gas.
 We have also found an amusing falling behavior of entropy-per-volume
 versus the $\mu$-parameter, i.e. strength of deformation.
 These remarkable facts can be of importance for more detailed future
investigations of real Bose like gases, along with similar use of
the result on $T_c^{(p,q)}$ of $p,q$-Bose gas model.

\vspace{-1mm}

\acknowledgments This work was partly supported by the Special
Program of the Division of Physics and Astronomy of NAS of Ukraine,
and by the Grant (A.P.R.) for Young Scientists of the NAS of Ukraine
$No.$ 0113U004910.

\end{document}